\documentstyle[12pt,aaspp4]{article} 
\def\tempest%
{\begin{array}{ccc} 
1 & 1 & 1 \\ 
1 & 1 & 1 \\ 
4 & 3 & 8 
\end{array}}

\def\kms{{\rm km}\,{\rm s}^{-1}} 
 
\def\e{{\rm E}}
\def\rel{{\rm rel}}
\def\btheta{{\vec\theta}}
\def\bmu{{\vec\mu}}
\def\bpi{{\vec\pi}}
\def\Teff{{T_{\rm eff}}}
\begin{document}

\title{Applications of Microlensing to Stellar Astrophysics}
\author 
{Andrew Gould}
\affil{Department of Astronomy, Ohio State University, Columbus, 
OH 43210, USA}
\affil{Laboratoire de Physique Corpusculaire et Cosmologie,
Coll\` ege de France, 11 pl.\ Marcelin Berthelot, F-75231, Paris, France}
\affil{E-mail: gould@astronomy.ohio-state.edu} 
\begin{abstract} 

	Over the past decade, microlensing has developed into a powerful
tool to study stellar astrophysics, especially stellar atmospheres,
stellar masses, and binarity.  I review this progress.  
Stellar atmospheres can be
probed whenever the source in a microlensing event passes over the
caustic (contour of infinite magnification) induced by the lens because
the caustic effectively resolves the source.  Broad-band observations
of four events have yielded limb-darkening measurements, which in essence
map the atmospheric temperature as a function of depth.  And now, for
the first time, spectroscopic observations of one event promise
much richer diagnostics of the source atmosphere.  In the past two years,
a practical method has finally been developed to systematically measure
the lens masses in microlensing events.  This will permit a census of
all massive objects, both dark and luminous, in the Galactic bulge, including 
low-mass stars, brown dwarfs, white dwarfs, neutron stars, and black holes.
The method combines traditional ground-based photometry
with astrometric and photometric measurements by the 
{\it Space Interferometry Mission (SIM)} in solar orbit.  Using a related
technique {\it SIM} can also obtain accurate ($\la 1\%$) mass measurement
of a dozen or so nearby stars, thus enabling precision tests of stellar
models.  Binary lenses can give rise to dramatic and easily detectable
microlensing signatures, even for large mass ratios.  This makes microlensing
a potentially powerful probe of the companion mass distribution, especially
in the Galactic bulge where this function is difficult to probe by other 
techniques.

\end{abstract}

\section{Brief History} 

	While microlensing observations were originally proposed as a means
to search for dark matter in the form of massive compact
halo objects (Paczy\'nski 1986), and have proved very effective for
that purpose (Alcock et al.\ 1998,2000c; Lasserre et al.\ 2000), they
have also been directed from the outset toward the Galactic
bulge (Udalski et al.\ 1993) where the vast majority of events were expected
to be due to ordinary stars in the Galactic disk 
(Paczy\'nski 1991; Griest et al.\ 1991) and the bulge itself
(Kiraga \& Paczy\'nski 1994).  The event rate reported
by OGLE (Udalski et al.\ 1994) and MACHO (Alcock et al.\ 1997a) was 
substantially too high to be consistent with any axisymmetric model of
the Galaxy (Gould 1994c; Kuijken 1997), or even with any plausible
barred model (Binney, Bissantz, \& Gerhard 2000).  Hence, the importance
of microlensing as a probe of Galactic structure was recognized almost from
the beginning, leading the EROS collaboration (Derue et al.\ 1999) to
extend their surveys from the Galactic bulge to the spiral arms.

	By contrast, the subject of the present review (microlensing
applications to stellar astrophysics) was initially much
slower to develop.  It is true that Refsdal (1964) first proposed
using microlensing to measure stellar masses more than 3 decades ago,
but the idea remained dormant until Paczy\'nski (1995) resurrected it,
and not a single candidate for such a measurement was identified in the
literature until last year (Salim \& Gould 2000).  There is no mention
in the literature that stellar atmospheres might lead to observable
signatures until Witt (1995), and the initial emphasis was
on using these effects to learn more about the lens
(Loeb \& Sasselov 1995; Gould \& Welch 1996), rather than about
the atmosphere of the source.   

	Despite its late start, this aspect of microlensing has witnessed
enormous progress over the past five years, partly because of intensifying
theoretical interest, but mainly because of the emergence of three groups
(MACHO/GMAN, MPS, PLANET) who dedicate themselves to the intensive 
microlensing follow-up observations that are required to probe these effects.
In this review, I first summarize the basics of microlensing (\S\ 2), and
then cover three major topics, stellar mass measurements (\S\ 3), binary
distribution functions (\S 5), and
resolution of stellar atmospheres (\S\ 6).  In the interval (\S\ 4), I 
give a brief introduction to binary microlensing.

\section{Microlensing Basics}

\begin{figure}
\epsfxsize \hsize
\epsfbox{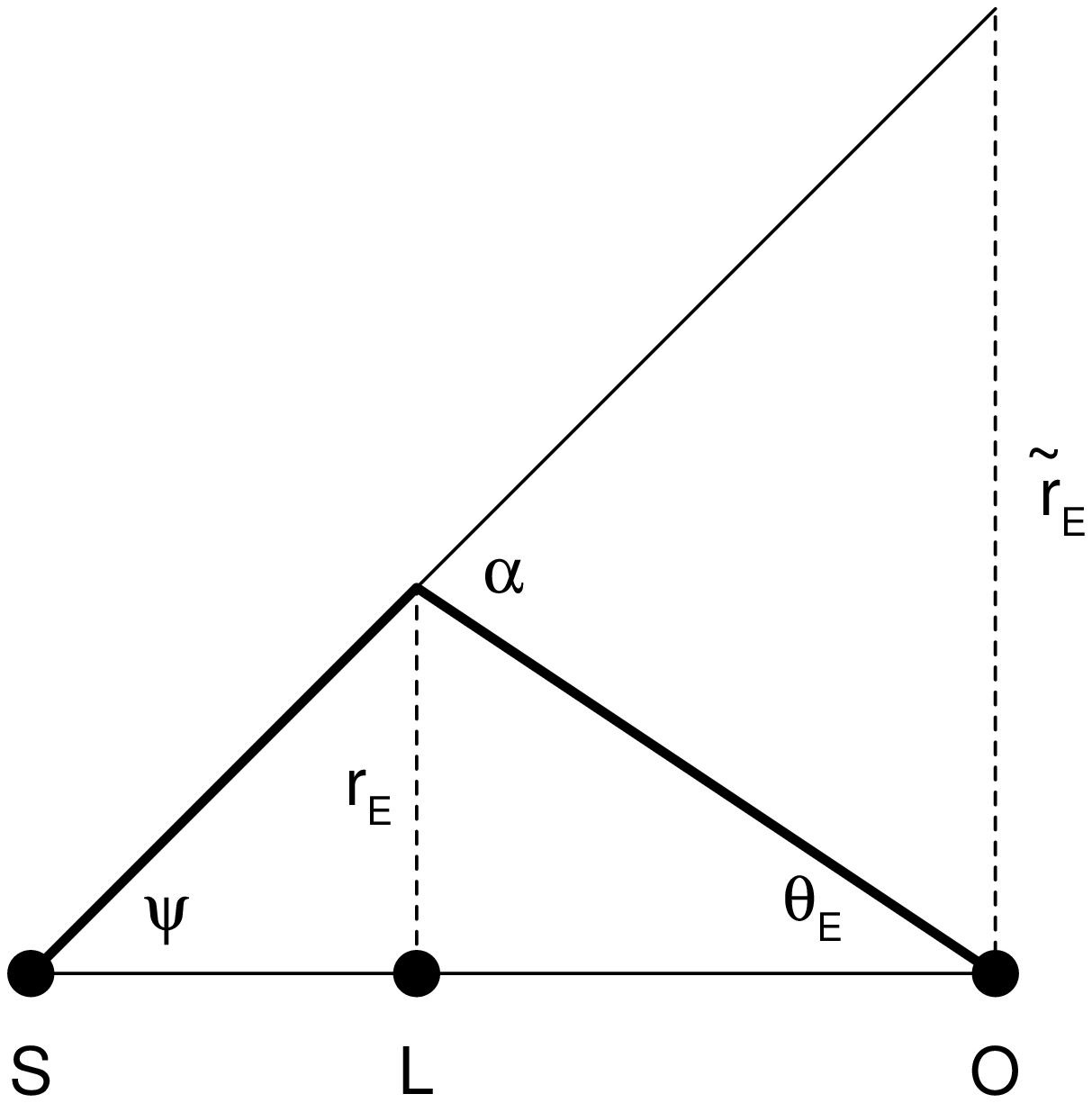}
\caption[junk]{\label{fig:one}
Microlensing geometry.  Bold curve shows the path of the light
from the source (S) to the observer (O) being deflected by the lens (L) 
of mass $M$.  The deflection angle is $\alpha=4GM/r_\e c^2$, where $r_\e$
is the Einstein radius shown as a dashed line.  The image is displaced
from the source by the angular Einstein radius $\theta_\e$.  The
Einstein radius projected onto the observer plane
is $\tilde r_\e$.  This diagram allows one to see immediately the relations 
between the
observables ($\theta_\e$, $\tilde r_\e$) and the physical parameters
$(M,\pi_\rel)$.  First, under the small-angle approximation, 
$\alpha/\tilde r_\e=\theta_\e/r_\e$, so 
$\tilde r_\e\theta_\e = \alpha r_\e= 4GM/c^2$.  Second, by the exterior-angle
theorem, $\theta_\e = \alpha - \psi = \tilde r_\e/D_L - \tilde r_\e/D_S$,
where $D_L$ and $D_S$ are the distances to the lens and source.  Hence,
$\theta_\e/\tilde r_\e = \pi_\rel/\rm AU$, where $\pi_\rel$ is the lens-source
relative parallax.  From Gould (2000c).  Copyright American Astronomical
Society, reproduced with permission.
}
\end{figure}

	Microlensing occurs when a massive object (``the lens'') becomes
closely aligned with a more distant luminous object (``the source'').
General relativity predicts that a lens of mass $M$ will deflect the light 
from the source by an angle $\alpha = 4 G M/b c^2$, where $b$ is the
impact parameter of the light trajectory relative to $M$.  This formula has 
been verified by Hipparcos measurements to be accurate to within 0.3\% 
(Froeschle, Mignard \& Arenou 1997).  If the lens is a point mass (or more
generally, spherical) and the lens and source are perfectly aligned with
the observer, then there is axially symmetry, and the source is imaged
into a ring of angular radius $\theta_\e$ (Einstein 1936), called the
``angular Einstein ring''.  Its projection onto the plane of the observer is
called the ``projected Einstein ring'', $\tilde r_\e$.  From simple 
geometric considerations (see Fig.\ \ref{fig:one}), one immediately finds
\begin{equation}
\tilde r_\e\theta_\e = {4 G M\over c^2},\qquad
{\theta_\e\over \tilde r_\e} = {\pi_\rel\over \rm AU},
\label{eqn:geomeqs}
\end{equation}
where $\pi_\rel$ is the lens-source relative parallax.   These
are easily solved,
$$
\theta_\e = \sqrt{\kappa M \pi_\rel},\quad
\pi_\e\equiv{{\rm AU}\over \tilde r_\e} = \sqrt{\pi_\rel\over \kappa M},
$$
\begin{equation}
\kappa\equiv {4 G\over c^2{\rm AU}}\simeq {8\,{\rm mas}\over M_\odot},
\label{eqn:geomsolv}
\end{equation}
where the ``microlens parallax'' $\pi_\e$ contains the same information
as $\tilde r_\e$ but in a more convenient form.  

	In the more general case, the source is not perfectly aligned with
the lens, so the axial symmetry is broken.  After a small bit of algebra,
one finds that the angular separation between the source and the lens 
$(\theta_\rel=\theta_s -\theta_l)$ 
and the angular separation between the image and the lens
$(\theta_I)$ are related by
\begin{equation}
\theta_I^2 - \theta_I\theta_\rel = \theta_\e^2,\quad
{\theta_{I\pm}\over\theta_\e} = {u\pm\sqrt{u^2+4}\over 2},
\quad u\equiv {\theta_\rel\over \theta_\e},
\label{eqn:thetaieq}
\end{equation}
which implies that there are two images, one on either side of the lens.  
Since for typical bulge events $\pi_\rel\sim 0.04\,\rm mas$,
the image separations,
$\sim 2\theta_\e\la 1\,$mas, are far smaller than can be resolved with
any existing instrument.  Hence, the only microlensing effect that has
been observed to date is the magnification, $A$.  By Liouville's theorem,
surface brightness is conserved, so $A$ is equal to the
ratio of the area of the image to that of the source.  For small sources,
this is given by the Jacobian of the transformation implied by equation
(\ref{eqn:thetaieq}).  Combining both image magnifications, $A=A_+ + A_-$, 
one finds
\begin{equation}
A_\pm = {A\pm 1\over 2},\quad A= {u^2 + 2\over u\sqrt{u^2+4}}.
\label{eqn:adef}
\end{equation}
If the observer, source, and lens are all in rectilinear motion, then
$u$ varies according to the Pythagorean theorem
\begin{equation}
u = \sqrt{u_0^2 + {(t-t_0)^2\over t_\e^2}},\quad 
t_\e\equiv {\theta_\e\over\mu_\rel},
\label{eqn:udef}
\end{equation}
where $t_0$ is the time of closest approach, $u_0= u(t_0)$, 
$t_\e$ is the ``Einstein timescale'',
and $\mu_\rel$ is the amplitude of
the lens-source relative proper motion.  Thus, three
parameters determine a standard microlensing event, $t_0$, $u_0$, and
$t_\e$.  This is both a blessing and a curse: a blessing because the
simplicity of microlensing light curves allows them to be robustly
distinguished from other much more common forms of stellar variability, and
a curse because only three parameters can be recovered from a normal
microlensing event.  Moreover, the only one of these parameters that
carries any information about the lens,
\begin{equation}
t_\e={\sqrt{\kappa M\pi_\rel}\over\mu_\rel},
\label{eqn:tedef}
\end{equation}
is a complicated combination of
the lens mass, and the lens and source distances and transverse velocities.

\section{Masses of Microlenses}

\subsection{Bulge Lenses}

	Thus it was always recognized that microlensing surveys would not
yield mass measurements for individual events.  The best one could hope
for would be statistical statements based on 
the observed distribution of timescales, equation (\ref{eqn:tedef}),
and assumptions about the 
underlying distributions of $\pi_\rel$ and $\mu_\rel$
(Mao \& Paczy\'nski 1996).  Early efforts
to apply this approach to bulge events by Zhao, Spergel, \& Rich (1995) 
and Han \& Gould (1996) revealed two types of problems.  First, there
were an excess of short events $(t_\e\la 10\,$days) relative to what one 
would expect if the bulge mass function (MF) were similar the local disk MF 
measured from {\it HST} star counts (Gould, Bahcall, \& Flynn 1997), and second
there was also an excess of long events ($t_\e\ga 50\,$days).  Alternatively,
one could characterize the situation as problems in the normalization 
and shape of the timescale distribution:  there are too many events
overall, and too many in the wings relative to the center.  A later
and more thorough analysis by Peale (1998) continued to show a strong
excess of short events over what could be expected from lower main-sequence
stars if the MF were similar to the local one, although Peale (1999)
argued that if the bulge contained a population of brown dwarfs similar
to the local one discovered by 2MASS, there would actually be a deficit
of short events.

	Han (1997)
argued that much of the short-event excess could be explained as due
to events with intrinsically faint sources which found their way into
the surveys by ``amplification bias'', and whose timescales were 
consequently being systematically underestimated.  Zoccali et al.\ (2000)
used NICMOS on {\it HST} to measure the bulge MF down to $0.15\,M_\odot$
and found that it indeed contains far more low-mass stars than the local
MF of Gould et al.\ (1997), and therefore should give rise to more 
short events.  

	Nevertheless, when Alcock et al.\ (2000b) applied a more sophisticated
image differencing analysis to MACHO bulge data, which should have removed
the effects amplification bias, the twin inconsistencies (normalization and
shape) between the observed and predicted timescale distributions remained
basically intact, albeit at a reduced level.  Moreover, an analysis of a 
largely independent data set of MACHO bulge clump giants by Popowski et al.\
(2001) confirms both the excess of long events and the resulting
high optical depth that Binney et al.\ (2000) found so difficult to
reconcile with models.

\begin{figure}
\epsfxsize \hsize
\epsfbox{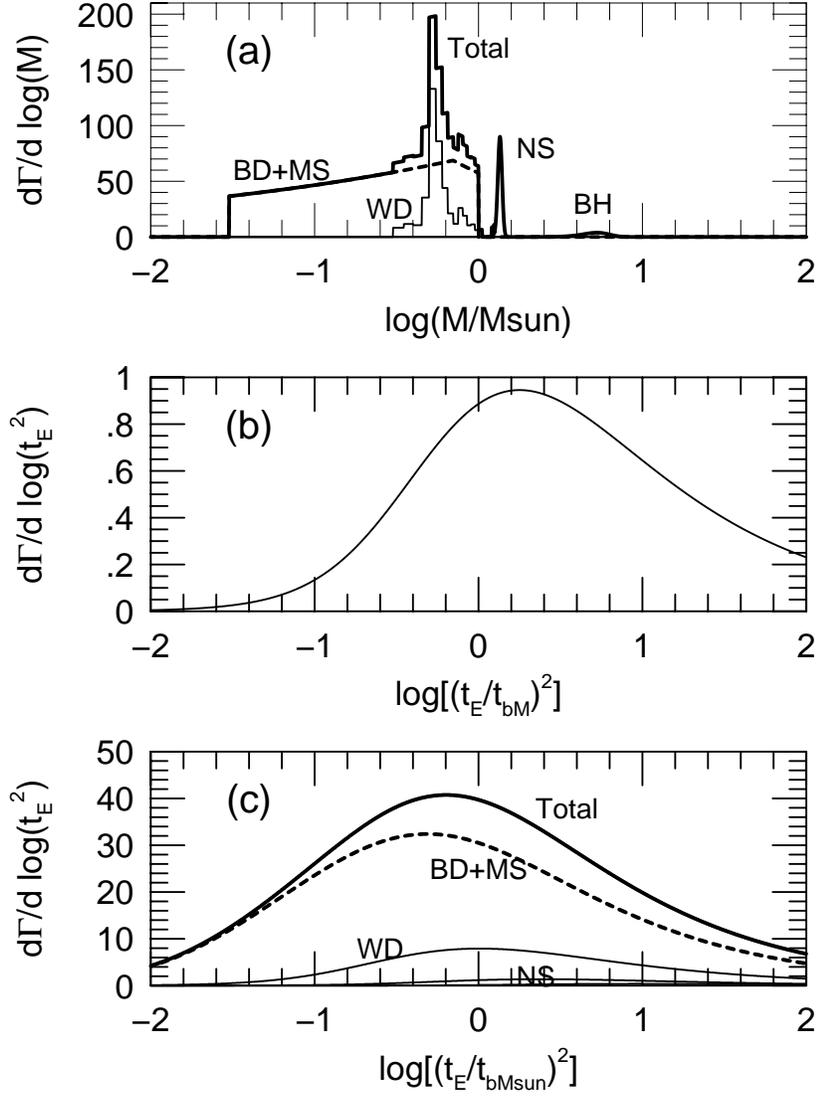}
\caption{\label{fig:two}
Rates of microlensing events toward the bulge by mass (panel {\it a}) and 
time scale (panel {\it c}) for
MS+BDs ($0.03\,M_\odot<M<1\,M_\odot$) 
({\it bold dashed curve}) 
and WD, NS, and BH remnants ({\it solid curves}).  The
total is shown by a {\it bold solid curve}.   The mass model ({\it a}) 
is described in \S\ 2 of Gould (2000b).  
It is convolved with the time scale distribution 
at fixed mass ({\it b}) derived in \S\ 2.2 of Gould (2000b), 
to produce the observable time scale distribution ({\it c}).  
All three classes of remnants
are clearly identifiable in the mass distribution which could be extracted
from SIM observations, but are utterly lost
in the time scale distribution.  The normalizations in panels
({\it a}) and ({\it c}) are for 100 events.  Panel ({\it b}) is
normalized to unity.  From Gould (2000b).  Copyright American Astronomical
Society, reproduced with permission.
}
\end{figure}

	However, as shown in Figure \ref{fig:two}, there is a fundamental
limit to how much information about the MF can be extracted from 
microlensing timescales alone.  Panel (a) shows a plausible
bulge MF decomposed into main-sequence stars and brown dwarfs (MS+BDs),
white dwarfs (WDs), neutron stars (NSs), and black holes (BHs).  The
MS MF in the range $0.15\,M_\odot < M< 0.9\,M_\odot$ is taken from actual 
measurements by Zoccali et al.\ (2000), but the other components are based on
Gould's (2000b) conjectures.  In particular, the cut off in the BD MF
at $0.03\,M_\odot$ is fairly arbitary.  Panel (b) shows the distribution
of timescales expected for microlensing events of fixed lens mass, $M=M_\odot$,
towards a field at projected distance, $b$, from the Galactic center and 
assuming an isotropic bulge velocity dispersion, $\sigma$.
The normalization of the timescale distribution is
$t_{b M_\odot}\equiv (2 G M_\odot/b)^{1/2}/\sigma c$.  The timescale 
distribution is shown as a function of $t_\e^2$ rather than $t_\e$ to
make it directly comparable to the MF, since for fixed $\pi_\rel$
and $\mu_\rel$, $M\propto t_\e^2$ (see eq.\ \ref{eqn:tedef}).  Note
that the FWHM of this distribution is a factor $\sim 100$.  Since this
is larger than the full extent of the MF in Panel (a), it follows that one can
learn very little about the MF beyond its mean and variance from timescale
observations alone, even assuming that the bulge velocity distribution
and ``contamination'' from foreground disk lenses were understood perfectly.
This conclusion is illustrated in Panel (c), which gives the predicted 
distribution of timescales formed by convolving Panels (a) and (b).  Note 
that all of the sharp features in Panel (a) are utterly obliterated, so
that it is impossible to pick out BDs, WDs, NSs, or BHs individually or
even statistically.  Thus, although finding the mean and variance of the
MF would be very important, and in particular would provide the only clue
we have as to the BD cutoff in the bulge,  all of the detailed information
about the large numbers (several hundred to date) of dark 
(BD, WD, NS, BH) lenses being detected toward the bulge would be lost.

	The solution is to find the lens masses and distances for
individual events.  Both $M$ and $\pi_\rel$ can be determined if
$\theta_\e$ and $\tilde r_\e$ are measured  (see eq.\
\ref{eqn:geomeqs}).  Fortunately, these two quantities are both
``observables'': $\theta_\e$ can be measured if it can be compared to
some ``standard angular ruler'' in the plane of the sky, and $\tilde r_\e$
can be measured if it can be compared to some ``standard physical ruler''
in the plane of the observer.  See my earlier review (Gould 1996)
for the large number of ideas to measure $\theta_\e$ and
$\tilde r_\e$ in various circumstances.  A few new ideas have been advanced
since then (Han \& Gould 1997; Hardy \& Walker 1995; Gould \& Andronov 1999;
Honma 1999).
  Unfortunately, to date there have been only 
about a half dozen measurements
each of $\theta_\e$ (Alcock et al.\ 1997b,2000a; 
Albrow et al.\ 1999a,2000,2001a; 
Afonso et al.\ 2000)
and of $\tilde r_\e$ (Alcock et al.\ 1995; Bennett et al.\ 1997; Mao 1999,
Soszy\'nski et al.\ 2001).
In no case have both been measured for the same event,
so that to date there is not a single lens mass measurement.
The problem is that the two standard rulers that have been applied,
the angular size of the source (Gould 1994a; Nemiroff \& Wickramasinghe 1994;
Witt \& Mao 1994) which is known from its dereddened color and magnitude
and the color/surface-brightness relation (van Belle 1999), and the
physical size of the Earth's orbit (Gould 1992), are available only for
very special, almost non-intersecting, subclasses of events.  These are
respectively, caustic-crossing and very long ($t_\e\ga 90\,$days) events.

	However, work over the last five years has developed a practical
method to obtain both $\theta_\e$ and $\tilde r_\e$ for a large and
representative sample of events.  First, 
Hog, Novikov \& Polnarev (1995), Walker (1995), and Miyamoto \& Yoshii (1995)
showed that $\theta_\e$ can in principle be determined 
from precision measurements of the {\it centroid} of the two lensed images,
$\btheta_c \equiv (A_+\btheta_{I+}+A_-\btheta_{I-})/A$. 
Boden, Shao, \& Van Buren (1998) and Paczy\'nski (1998) then showed that
the proposed {\it Space Interferometry Mission (SIM)} would be capable
of making such measurements.  From equations 
(\ref{eqn:thetaieq}) and (\ref{eqn:adef}), the astrometric shift
$\delta\btheta_c$ of the centroid relative to the position of the source
in the absence of lensing is given by,
\begin{equation}
\delta\btheta_c\equiv\btheta_c - \btheta_\rel = 
{\btheta_\rel\over u^2 + 2}\,.
\label{eqn:ellipse}
\end{equation}
Although it is not immediately obvious from equation (\ref{eqn:ellipse}),
if $\btheta_\rel$ moves in a straight line, then $\delta\btheta_c$
traces an ellipse, and its maximum amplitude (at $u=2^{1/2}$) is
$\theta_\e/8^{1/2}$.  Recall from the discussion following 
equation (\ref{eqn:thetaieq}) that $2\theta_\e\la 1\,$mas, so that the
two images cannot be resolved.  However, it is much easier to centroid
an image than resolve it, and in particular {\it SIM} is expected to
reach an astrometric precision of $\sim 4\,\mu$as.  Miyamoto \& Yoshii (1995),
Boden et al.\ (1998),
and Paczy\'nski (1998) also noted that the motion of the Earth would
cause $\btheta_\rel$ to deviate from a straight line, and so induce distortions
on the ellipse, in principle permitting the measurement of $\tilde r_\e$, and
so of $M$, using astrometry alone.
However, it turns out that these parallax distortions are unmeasurably small
in most cases, as shown both analytically and numerically by Gould \& Salim 
(1999).

	Nevertheless, simultaneous measurement of $\tilde r_\e$ and
$\theta_\e$ should be possible for a large number of events using {\it SIM}.
 From equation (\ref{eqn:geomsolv}), and recalling that $\pi_\rel\sim 0.04$,
it follows that 
$\tilde r_\e\la 10\,$AU.  Hence, the event will have substantially different
parameters if viewed from a satellite in Earth-trailing orbit
$(t_{0,\rm sat},u_{0,\rm sat},t_{\e,\rm sat})$
than it does from the Earth
$(t_{0,\oplus},u_{0,\oplus},t_{\e,\oplus})$.  One can then determine
$\bpi_\e$ up to a four-fold degeneracy (Refsdal 1966),
\begin{equation}
\bpi_\e = {\rm AU\over |{\bf D}_{\rm sat}|}
\biggl({\Delta t_0\over t_\e},\Delta u_0\biggr),\qquad
\Delta t_0\equiv t_{0,\rm sat}-t_{0,\oplus},\quad
\Delta u_0\equiv \pm|u_{0,\rm sat}\pm u_{0,\oplus}|,
\label{eqn:pieeval}
\end{equation}
where ${\bf D}_{\rm sat}$ is the Earth-Satellite separation vector
projected onto the plane of the sky, and the direction of $\bpi_\e$ is
taken to be that of the lens-source relative proper motion, $\bmu_\rel$,
with ${\bf D}_{\rm sat}$ defining the $x$-axis.  See Figure 5 from my
previous review (Gould 1996).

	The four-fold degeneracy arises because one does not know
on which side the source passes the lens and hence whether
$u_{0,\rm sat}$ and $u_{0,\oplus}$ should be regarded effectively as
``positive'' or ``negative''.  See Figures 2 and 3 from Gould (1994b).
However, this degeneracy can usually be resolved 
by measuring the small difference,
$\Delta t_\e=(t_{\e,\rm sat} - t_{\e,\oplus})$,
which is proportional to $\Delta u_0$
(Gould 1995; Boutreux \& Gould 1996; Gaudi \& Gould 1997a).  

	Gould \& Salim (1999) pointed out that since {\it SIM} does
astrometry by {\it counting} photons as a function of fringe position,
it can simultaneously do photometry and hence can (in conjunction with
ground-based photometry) measure $\bpi_\e$.  Moreover, since the axis of
the astrometric ellipse described by equation (\ref{eqn:ellipse}) is
parallel to $\bmu_\rel = -d\btheta_\rel/dt$, {\rm SIM} astrometry provides
a second method to distinguish among the four solutions given by
equation (\ref{eqn:pieeval}), which each would imply different directions
for $\bpi_\e$ (and hence $\bmu_\rel$).
Finally, since {\it SIM} automatically measures $\pi_s$ and $\bmu_s$,
the parallax and proper motion of the source, it can also determine
$\pi_l = \pi_\rel + \pi_s$ and $\bmu_l = \bmu_\rel + \bmu_s$, the
parallax and proper motion of the lens.  Salim \& Gould (2000) showed that
for bright $(I\sim 15)$ sources, {\it SIM} could measure $M$ 
accurate to $\sim 5\%$ in 5 hours of observation time, which is approximately 
the resolution of the mass function illustrated in Figure \ref{fig:two}.

	Han \& Kim (2000) 
have proposed another method to measure $\tilde r_\e$ by
comparing {\it SIM} {\it astrometry} to that of ground-based interferometers.
The principle is broadly similar to the above {\it photometric} comparison,
but in this case there is no degeneracy.

	Such a MF measurement would automatically yield substantial 
information about the rate of binarity and the distribution of binary mass
ratios.  Although a large fraction of stars are believed to be in binaries,
for a binary to be recognizable from a microlensing light curve, its
projected separation must be close to $\theta_\e$ (see \S\ 4).  As a result,
only $\sim 5$--$10\%$ of 
microlensing events are photometrically distinguishable 
from point lenses.  However, in a series of paper, C.\ Han and his 
collaborators
have demonstrated that a much larger fraction of binaries can be detected
and accurately characterized astrometrically (Chang \& Han 1999; Han, Chun
\& Chang 1999; Gould \& Han 2000; Han 2001).

\subsection{Nearby Lenses}

	Refsdal (1964) showed that it would be possible
to determine the mass of a nearby star by measuring its deflection of light
from a more distant field star.  While the mathematics for this type
of microlensing are formally identical to those described in \S\ 3.1, the
physical conditions, observational requirements, and scientific motivations
all differ radically.  Because the astrometric microlensing effect falls off 
only as $u_0^{-1}$ (compared to $u_0^{-4}$ for the photometric effect)
the encounters typically have $u_0\gg 1$.  Hence, the photometric
effect is negligible, and only the astrometric
effect survives (Miralda-Escud\'e 1996).  In this regime, the source appears 
displaced by $\alpha' = \kappa M\pi_\rel/\theta_\rel$, and the ellipsoidal
path of deviation becomes circular.  
Since the lens and source
are both luminous, their relative separation and parallax,
$\theta_\rel$ and $\pi_\rel$ are measurable astrometrically,
and hence $M$ can be inferred directly from the measurement of $\alpha'$
(together with a time series of measurements to determine $\pi_\rel$
and $\theta_\rel$).
Second, the motivation for doing these observations 
is not the cataloguing of dark objects, but the precision measurement 
of the mass of luminous ones.  This is the only practical
method to obtain accurate
masses for single stars (except the Sun), and as we shall
see below, the method is strongly biased toward metal-poor stars 
(because of their high proper motions) for which
there are at present no reliable mass measurements at all. Thus, it
is complementary to the standard techniques for measuring stellar masses
using visual and eclipsing binaries (e.g.\ Henry \& McCarthy 1993).

	This idea remained dormant for 30 years until Paczy\'nski (1995)
resurrected it. It was further studied in the context of the
accurate measurements possible using {\it SIM} and {\it Global Astrometric
Interferometry for Astrophysics  (GAIA})
by Miralda-Escud\'e (1996), Paczy\'nski (1998), and Dominik \& Sahu (2000).
Given an astrometric instrument of sufficient precision,
the central problem is the identification of lens-source pairs that will
come close enough to permit an accurate mass measurement.  
In principle, one would like to consider all possible pairs of stars on 
the sky, but this is not possible with existing catalogs because these lack
proper motions for most stars.  Hence a more focused approach is required.

    For a fixed amount of observing time to be scheduled during a fixed project
lifetime, the probability that a given nearby star can have its mass measured
to a given fractional precision is approximately, 
\begin{equation}
P \propto \pi_l\mu_l M N_s,
\label{eqn:probnearby}
\end{equation}
where $N_s$ is the density of sources behind the lens.  Hence,  nearby,
high proper motion stars close to the Galactic plane have the best chance.
Since nearby stars also tend to have high proper motions, Paczy\'nski (1998)
advocated checking the paths of proper-motion stars for future encounters
with background sources, and specifically estimated that {\it Hipparcos}
catalog stars should have dozens of such encounters.  Sahu et al.\ (1998)
predicted three encounters by tracking the future paths of 500 high proper
motion WDs on archival sky survey plates complemented with ground-based
observations, although they did not identify these explicitly.

	Gould (2000a) and Salim \& Gould (2000) outlined a three step 
procedure to systematically find 
candidates by combining a proper-motion catalog, e.g.\ {\it Hipparcos} or
Luyten (1979,1980, hereafter NLTT) with an all-sky position
catalog, e.g., USNO-A (Monet et al.\ 1998).  First, estimate the lens
distance using parallax for {\it Hipparcos} stars or a reduced proper motion
diagram for NLTT stars.  Since the mass error scales as the distance squared,
the list of possible candidates would mushroom without this step.
Second, search in the neighborhood of the {\it future} path of these stars
for sources whose {\it archival} (e.g.\ 1950) USNO-A positions put them close
enough for a significant deflection.  Third, make follow-up observations of
the lens-source pairs to confirm their encounter parameters.  The third
step is required for three reasons: 1) the USNO-A positions are accurate
only to 250 mas, which can be a significant fraction of the impact parameter
in some cases, 2) the source stars will have moved due to their 
(unknown) proper motions,
which are generally expected to be of order 2 mas/yr, but could be larger,
3) the NLTT proper motions are accurate only to 20 mas/yr, which implies
an error in 2010 position of $1.\hskip-2pt ''2$.

	Salim \& Gould (2000) carried out this search through the second
step.  They found 11 encounters for {\it Hipparcos} stars during the interval
2005-2015 for which 1\% mass measurements could be obtained in less than
14 hours of {\it SIM} time.  The errors in 2010 positions due to reasons
(1) and (2) above were not expected to be large for these stars.  
Salim \& Gould (2000) also produced a table of 180 NLTT stars for follow-up
observations among which they expected $\sim 10$ will have mass errors 
comparable to the 11  {\it Hipparcos} stars.  The large errors in NLTT 
proper motions are responsible for this huge ratio of stars requiring follow-up
to those that will be found useful.  Salim \& Gould (2001) are undertaking
these follow-up observations.

	What improvements can be hoped for in the future?  These would
come mainly from rectifying four shortcomings in the present catalogs.
First, NLTT is nominally complete only for $V\la 18.5$, $\delta >-20^\circ$,
and $|b|>10^\circ$.  
Second, NLTT proper motions are accurate only to 20 mas/yr, implying
a barely acceptable error of $1.\hskip-2pt''2$ in the predicted position
of the encounter.  
Third, NLTT archival positions are accurate only to a few
arcseconds, making them essentially useless for predicting encounters.
This problem can be circumvented for $\delta>-15^\circ$ by identifying
NLTT stars on USNO-A, but
fourth, USNO-A is missing essentially esstentially all NLTT stars
for $\delta<-15^\circ$.  Fifth,
USNO-A lacks proper motions, so that encounters
with slow nearby stars cannot be accurately predicted.

	Some, but not all of these problems will be resolved with
the publication of either of two projected all-sky position
and proper motion surveys, USNO-B (D.\ Monet 1999, private communication) 
or Guide Star Catalog II (Baruffolo, Benacchio, \& Benfante 1999).  
These are expected to go a magnitude deeper than NLTT, to have accurate
positions and proper motions, and to cover the whole sky.  However,
the dearth of high-proper motion stars in southern catalogs derives
fundamentally from the lack of coeval 2-color surveys in this region of
the sky, which renders difficult their identification.   Since no
additional surveys are planned, this problem may persist.

	In summary, with some additional work, one can expect perhaps a 
few dozen encounters per decade.

\section{Binary Microlensing}

	The other applications that I describe all make use of binary
microlenses, i.e., microlensing where the mass distribution is composed
of two point masses.  	Binary microlensing is one of the most active fields 
in microlensing today.  In part this is due to the mathematical
complexity of the subject and in part to the demands that are being placed
on theory by new, very precise observations of binary events.  
Chang (1981) made the first study of binary microlenses in her thesis, which
included a detailed investigation of the important limiting case
of a high-mass ratio (``planetary'') binary.  See also Chang \& Refsdal
(1979,1984).  Schneider
\& Weiss (1986) made a comprehensive study of binary lenses despite the
fact that they never expected any to be detected (P.\ Schneider 1994,
private communication), in order to learn about caustics in quasar
macrolensing.  
Indeed caustics are the main new features of binaries
relative to point lenses.  These are closed curves in the source plane
where a point source is infinitely magnified.  The curves are composed of
3 or more concave segments that meet at cusps.  Binary lenses can have
1, 2, or 3 closed caustic curves.  If the two masses are separated by 
approximately an Einstein radius, then there is a single 6-cusp caustic.
If they are separated by much more than an Einstein ring, then there are
two 4-cusp caustics, one associated with each member of the binary.  If the
masses are much closer than an Einstein ring, there is a central 4-cusp
caustic and two outlying 3-cusp caustics.  Figure 3 shows two cases of
the 6-cusp caustic, one close to breaking up into the two caustics 
characteristic of a wide binary and the other close to
breaking up into the three caustics characteristic of a close binary.  
See also Schneider, Ehlers \& Falco (1992).  Witt (1990) developed a simple
algorithm for finding these caustics.  Multiple-lens systems can have
even more complicated caustic structures (Rhie 1997; Gaudi, Naber, \& Sackett
1998).

\subsection{Binary Lens Parameters}

	Recall from equation (\ref{eqn:udef})
that a point-lens light curve is defined by just three
parameters, $t_0$, $u_0$, and $t_{\rm E}$.  These three generalize to
the case of binaries as follows: $u_0$ is now the smallest separation of
the source relative to the center of mass (alternatively geometric center)
of the binary, $t_0$ is the time when $u=u_0$, and $t_{\rm E}$ is the
timescale associated with the combined mass of the binary.  At least three
additional parameters are required to describe a binary lens: the angle
$\alpha$ at which the source crosses the binary axis, the binary mass ratio
$q$, and the projected separation, $d$, 
of the binary in units of the Einstein ring.
Several additional parameters may be required in particular cases. If 
caustic crossings are observed, then the infinite magnification of the caustic
is smeared out by the finite size of the source, so one must specify
$\rho_*=\theta_*/\theta_{\rm E}$, where $\theta_*$ is the angular size of
the source.  If the observations of the crossing are sufficiently precise,
one must specify one or more limb-darkening coefficients for each band
of observation (see \S\ 6.1).
Finally, it is possible that the binary's rotation is detectable
in which case one or more parameters are required to describe it
(Dominik 1998; Albrow et al.\ 2000).  In addition, binary light curves often
have data from several observatories in which case one needs two parameters
(source flux and background flux) for each observatory.  

\subsection{Generic Caustic Crossings}

	Most stars are believed to be in binaries, but only of order 5--10\%
of microlensing events show recognizable signatures of binarity 
(e.g.\ Alcock et al.\ 2000a).  The reason is simple: binaries span about
7 decades in semi-major axis, but unless their projected separation is
within a factor $\sim 3$ of $\theta_\e$, the caustics are extremely small
and the magnification patterns closely resemble those of isolated lenses
(e.g. Di Stefano \& Mao 1996; Gaudi \& Gould 1997b).
As a result, most detected binaries are drawn from the relatively small
subclass with caustics whose dimensions are of order $\theta_\e$.
Since typically, $10^{-3}\la \rho_* \la 10^{-2}$, this implies that
the source is generally several orders of magnitude smaller than the caustic,
so that the caustic crossing usually takes place well away from any cusps.

A source inside 
a caustic will be imaged into five images, while outside the caustics it will
be imaged into three images.  Hence, at the caustic two images appear or
disappear.  These images are infinitely magnified.  In the immediate 
neighborhood of a caustic (assuming one is not near a cusp), the magnification
of the two new images diverges as $A_2 \propto (-\Delta u_\perp)^{-1/2}$,
where $\Delta u_\perp$ is the perpendicular separation of the source from
the caustic (in units of $\theta_\e$).  On the other hand, the
three other images are unaffected by the approach of the caustic, so
$A_3\sim$ const.  Hence, the total magnification is given by 
(Schneider \& Weiss 1987)
\begin{equation}
A = A_2 + A_3 \simeq \biggl(-{\Delta u_\perp\over u_r}\biggr)^{-1/2} 
\Theta(-\Delta u_\perp) + A_{cc},
\label{eqn:a23}
\end{equation}
where $u_r$ is a constant that characterizes the approach to the caustic,
$A_{cc}$ is the magnification just outside the caustic crossing, and
$\Theta$ is a step function.  For a source of uniform brightness, or
limb darkened in some specified way, one can therefore write a relatively
simple expression for the total magnification as a function $\Delta u_\perp$
(Albrow et al.\ 1999b; Afonso et al.\ 2000).
	
\section{Binary Companion Distribution}

Microlensing can potentially probe the distribution of binary companions 
of bulge stars as a function of mass ratio and, to a certain extent, 
separation. 
These binary distribution functions provide one of the major observational
constraints on theories of star formation.  Since the Galactic
bulge is the nearest elliptical/bulge type structure, and since these are
thought to contain the majority of stars in the universe, it is of
exceptional importance to understand the distribution of binaries in
this population.

	I mentioned in \S\ 3.1 that SIM astrometry would automatically yield
substantial information about bulge binaries.  However, a lot of work can
already be done today using ground-based photometry.
Alcock et al.\ (2000a) have conducted the only systematic search for binarity
to date.  Their study reveals both the promise and the pitfalls of this
technique.  On the one hand, caustic-crossing binaries yield an unambiguous
signature, and microlensing is sensitive to companions of very small mass.  
On the other hand,
there are a large number non-caustic crossers that are either not
recognizable at all as binaries (see \S\ 4.2) or whose binary parameters
are poorly determined.  Of course, one could adopt the approach of simply
excluding these from the sample (and modeling the selection function 
accordingly) but Afonso et al.\ (2000) showed that for one caustic
crossing binary with extremely good light-curve coverage, it was not
possible to definitively distinguish between two sets of binary parameters,
one where the companion was much heavier than the main perturber and separated 
from it by much more than an Einstein radius, and other where the companion was
lighter than the main perturber and closer to it than an Einstein radius.  
At about the same time,
Dominik (1999) showed that such wide/close degeneracies were generic
to binary microlensing, although this degeneracy does appear to be 
breakable in many individual cases (e.g., Albrow et al.\ 2001a).  Hence,
careful modeling will be required to go from microlensing detections 
of binaries to a mass-ratio distribution.

	Another problem (see \S\ 4.2) is that photometric 
microlensing is mainly sensitive to binaries only over about a 
decade of projected separation: outside this range the great majority of
binary events are indistinguishable from those due to a single lens.  
By searching for relatively rare events, this range can be extended only about 
another decade (Di Stefano \& Mao 1996; Gaudi \& Gould 1997b),
compared to the $\sim 7$ decades that binaries are known
to populate (Duquennoy \& Mayor 1991).   Nevertheless, microlensing could be 
combined with a variety of other techniques to probe all but about a decade 
in separations of bulge microlenses (Gould 2000b).  Microlensing
would be most sensitive to low-mass companions while other methods
would provide most of the information about the separation distribution.

\section{Stellar Atmospheres}

	The Sun appears brighter and bluer at its center than at its limb
because the surface of last scattering lies deeper in the Sun where
the atmosphere is hotter.  Hence, by measuring the solar profile in
various broad bands or spectral lines, one can learn about the atmosphere
as a function of height.  However, it has proven extremely difficult to make 
similar measurements for any star except the Sun, simply because they are
unresolved or, at best, barely resolved.  Caustic-crossing microlensing
events permit such resolution because, as the caustic passes over the
face of the star, different sections are strongly magnified at different
times.

\subsection{Limb-Darkening Measurements}

At one time, it was thought that broad-band profiles, i.e., 
limb darkening (LD), 
could be measured from eclipsing binaries (e.g.\ Wilson \& Devinney 1971;
Twigg \& Rafert 1980),
but Popper (1984) showed that the LD coefficients derived in this manner
were too large to be of use due to degeneracies with other parameters.
In contrast to stellar eclipses, planetary transits such as
HD 209458 can yield accurate LD measurements (Jha et al.\ 2000;
Deeg, Garrido \& Claret 2001).  However, apart from the Sun and
HD 209458 (both G dwarfs), and from the four
microlensing measurements described below, there has been only one modern
published LD measurement (Burns et al.\ 1997), which was of Betelgeuse and was 
carried out by means of interferometry.  The paucity of LD measurements
(as opposed to detections that could in principle be used to make 
measurements) may be due in 
part to an underappreciation of their importance.  I will return to this
point below.

	Witt (1995), Valls-Gabaud (1995), and Bogdanov \& Cherepashchuk (1995)
showed that microlensing light curves could
be affected by LD, but early papers on this subject
(e.g., Loeb \& Sasselov 1995; Gould \& Welch 1996) were primarily concerned 
with using this effect to learn more about the lens rather than the source.
Moreover, theoretical analysis was initially focused on source resolution
by point-mass lenses whereas, as we shall see below, all four measurements
made to date use binary microlenses.  Gaudi \& Gould (1999) have studied the
signal-to-noise properties of light curves resulting from source
transits of both point-lens caustics and binary fold caustics.  Rhie \& Bennett
(1999) have systematically investigated the observational requirements for 
accurately measuring LD parameters from fold-caustic crossings
for a range of parameterizations.

	Alcock et al.\ (1997b, the MACHO/GMAN collaboration) made the first 
microlensing detection of LD using the event MACHO 95-BLG-30, in which 
an M4 giant source ($\sim 60\,R_\odot$) was lensed by a point mass.
However, the modest
significance of the detection did not permit
a measurement of LD parameters.

	Albrow et al.\ (1999a, the PLANET Collaboration) made the first 
microlensing LD measurement
using the event MACHO 97-BLG-28\footnote{So named because it was the 28th
event alerted by the MACHO collaboration toward the Galactic bulge in 1997.
See http://darkstar.astro.washington.edu/},
whose source they found spectroscopically 
to be a K2 giant.
Both the event and the analysis were spectacular\footnote{Since I am so 
enthusiastic about this work,
and since I am a co-author on many Albrow et al.\ papers, I should make
clear that I had absolutely no connection with this one.}, with the result
that this is the best LD measurement to date.  The event was extraordinary 
in that it had a cusp crossing, which is a priori very unlikely (see \S\ 4.2).
This makes the event both more difficult to monitor intensively, and more
interesting. Cusp crossings are more difficult because they usually occur with 
little or no warning, so the onset of the crossing must be recognized in 
real time.   In fact, both the PLANET and MACHO/GMAN collaborations
alerted on the crossing within hours of its start.  They are more interesting
because the needle-like geometry of cusps makes them more similar to 
the point-like caustics of point-masses than to ordinary (fold) caustic 
crossings of binaries.  Gaudi \& Gould (1999) showed that point caustics were
potentially much more sensitive probes of stellar stucture than fold caustics,
but argued that they were also much less likely to be observed.

	The resulting intensive and accurate photometry from 3 sites,
together with the needle geometry of the cusp, enabled Albrow et al.\ (1999a) 
to make 2-parameter LD models of the source in $V$ and $I$, 
the only 2-parameter measurement using microlensing to date.
They found overall excellent agreement with models of a K2 source by
van Hamme (1993) and D\'\i as-Cordovas, Claret \& Gim\'enez (1995).

	Afonso et al.\ (2000) made intensive observations of the 
binary fold-caustic event MACHO 98-SMC-1, primarily to determine whether
the lens was in the Galactic halo (and so a contributor to the dark matter)
or in the SMC itself.  They found the latter, and in the course of doing
so also obtained 1-parameter LD coefficients in $V$, $R$, and
$I$.  The
source was a metal-poor A dwarf.  It is difficult to see how LD measurements
could be made of such a star by any method except microlensing, since there
are few, if any, in the Galaxy.  Afonso et al.\ (2000) could not find
models with which to compare their results.

	Albrow et al.\ (2000) obtained 1-parameter LD coefficients in
$I$ band for the extremely complicated binary event MACHO 97-BLG-41
in which a cool K giant source $(T\sim 5000\,$K) crossed two disjoint 
caustics.  That
is, there were a total of 4 caustic crossings, including 3 fold caustics
and one cusp.  The main interest in this event is that it was the first
for which binary rotation was measured.  The caustic crossings were
either missed or poorly covered, primarily due to bad weather.  As a 
consequence, the LD measurement had rather large ($\sim 20\%$) errors,
so that while it was consistent with the models of
Claret, D\'\i az-Cordov\'es \& Gim\'enez (1995), it could not challenge
these models.

	Finally, Albrow et al.\ (2001a) obtained 1-parameter LD coefficients
in $V$ and $I$ bands of the fold-caustic crossing event 
OGLE 99-BUL-23\footnote{
http://www.astrouw.edu.pl/\~{ }ftp/ogle/ogle2/ews/bul-23.html
(Udalski, Kubiak \& Szyma\'{n}ski 1997)}.  Based on the source's
position in the color-magnitude diagram, they estimated it to be a
G/K subgiant ($T\sim 4800\,$K).

	Albrow et al.\ (2001a) developed for the first time the
methods needed to use microlensing LD measurements to distinguish
between competing models of stellar atmosphseres.  First, they made
a much more thorough investigation of the errors in the LD coefficients.
Whereas previous studies (Afonso et al.\ 2000; Albrow et al.\ 2001a) had
quoted LD errors derived by fitting the light curve at fixed lens parameters,
Albrow et al.\ (2001a) included the errors due to correlations with other
parameters, and found in particular that the largest contribution came
from correlation with the lens geometry $(d,q)$.  Second, they made their
comparison with published atmosphere models in the {\it 2-dimensional} 
$(V,I)$ plane, which allowed them to take account of correlations
between these parameters in both the measurements and the models.  Third,
they compared their results to several different competing models and so 
were able to make quantitative statements about which models were
favored and by how much\footnote{I note that
this analysis was almost entirely the work of a graduate student, Jin An.}.  
See Figure \ref{fig:four}.

	Unfortunately, the LD measurements of Albrow et al.\ (2001a) only
marginally discriminate between models.  However, it should be possible
in the case of future events to obtain smaller errors in the LD parameters
and then to use the methods of Albrow et al.\ (2001a) to say which models
are more correct.  A significant remaining obstacle to doing this is
that linear LD does not accurately represent stellar profiles, at least
those that are predicted in models (Orosz \& Hauschildt 2001).  Hence,
differences between the way the model is sampled theoretically and the way
the star is effectively sampled by microlensing, can introduce subtle
differences in the linear LD coefficient.  Problems of this sort are
probably the main reason that many authors prefer to compare their results
directly with the predictions of models, rather than give parameterized
measurements (e.g., Jha et al.\ 2000).  However, if LD measurements are
to be used to {\it discriminate among models} (and not just confirm their
general superiority over uniform sources), then the comparison must be
made in a ``space'' that is large enough to encompass several models and
allows these models to range over parameters that are only partly constrained,
such as temperature and surface gravity.  An $(n\times m)$-dimensional space 
defined by $n$ LD parameters in each of $m$ bands 
can perform exactly this function (e.g. Albrow et al.\
2001a).  While there may be other ways achieve this end, none have come
to my attention.  Hence, it is important to develop a better parameterization
than the conventional linear one, or its more general power-law extensions.

	Heyrovsk\'y (2001) has made a very important advance in this
direction with his suggestion to model stellar profiles as a linear
combination of basis functions drawn from a principle component analysis
(PCA) of an ensemble of models.  If the models are even approximately
correct, then the PCA analysis will, by construction, generate a more
accurate representation of the stellar profile than the traditional
LD decomposition.  In order to compare two different ensembles of stellar 
models, it will probably be necessary to extend Heyrovsk\'y's (2001)
original idea to make PCA analyses of each.

\subsection{Full Spectral/Spatial Resolution}

	Spectroscopic effects in microlensing events were first discussed
by Maoz \& Gould (1994) and Valls-Gabaud (1995).
Valls-Gabaud (1996,1998) modeled the convolution of point-lens 
microlensing magnification patterns with spatially resolved stellar spectra 
and argued that it
should be possible to reconstruct the 3-dimensional atmospheric profile
from a series of spectral measurements.  Heyrovsk\'y \& Loeb (1997) worked out
an efficient algorithm for carrying out such calculations, and
Heyrovsk\'y, Sasselov \& Loeb (1999) applied this method to make detailed
predictions of the spectra of a microlensed M giant ($T=3750\,$K), 
including both low ($R=500$) and high ($R=500,000$) resolution.
Most importantly, they focused attention on specific regions of the
spectrum, notably the Balmer lines and TiO bands, that would vary
relative to the continuum as the microlensing event progressed.  

	Unfortunately, all of this work proceeded under the assumption
that the source would be resolved by a point-mass lens, whereas in 
practice the overwhelming majority of spectroscopic observations will be
of binary-lens caustic crossings
(Gaudi \& Gould 1999).
One reason for this is that point-mass caustic crossings are intrinsically
rarer (see \S\ 4.2), but a much deeper problem is that they cannot be
reliably predicted.  Since the observations require large ($\ga 4\,$m) class
telescopes to which individuals do not generally have dedicated access,
it is essential that the caustic crossing be predicted in advance.  Once
a source has entered a binary caustic, it inevitably must exit.  Hence
one usually has several days to weeks to make general preparations to
observe the crossing, and because fold crossings are characterized by
an inverse square-root singularity (eq.\ \ref{eqn:a23}), they can usually 
be accurately predicted a day or more in advance.

Alcock et al.\ (1997b) acquired spectra of an M4 giant in the 
high-magnification event
MACHO 95-BLG-30 and saw changes in H$\alpha$ and TiO near 6700\AA\ that they
suggested were due to center-to-limb variations in the spectral lines.
Lennon et al.\ (1996) obtained three 30 min exposures during a caustic crossing
of a warm $(T=6100\,$K) dwarf star using the ESO NTT with 3.3 \AA\   
resolution.  Although the source was magnified by a factor $\sim 25$
at the time of the observations (converting the NTT effectively from a
3.6 m to an 18 m telescope), they were unable to discern any differences
in profile shapes for the three observations, and hence were not able to
use the caustic crossing to resolve the source.

	The only microlensing event to be clearly spectroscopically 
resolved to date was EROS BLG-2000-5, in which a K3 giant source traversed
a binary-lens caustic.  In fact, this required the coordinated efforts of
3 microlensing collaborations plus a number of unaffiliated individuals.
The event was initially alerted by the EROS
collaboration\footnote{httP://www-dapnia.cea.fr/Spp/Experiences/EROS/alertes.html} in April 1999.  On 8 June, the MPS 
collaboration\footnote{http://bustard.phys.nd.edu/MPS/index.html} issued 
an anomaly alert saying the magnification had jumped, and this alert enabled 
the PLANET collaboration\footnote{http://thales.astro.rug.nl/\~{}planet/}
to obtain dense coverage of the first crossing.  Because first crossings
are not usually predictable, such coverage is extremely rare.  Using their
precise characterization of the first crossing as well as their detailed
measurements of the intra-caustic light curve, PLANET was able to reliably
predict not only the time but also the {\it duration} of the second crossing, 
which latter was an unprecedentedly long 4 days.  The long second crossing
meant that the spectra should be taken on successive nights, and so made
possible the use of northern as well as southern telescopes.  In the
end, low-resolution ($R\sim 1000$) spectra were taken from the VLT on four 
successive nights (Albrow et al.\ 2001b) 
and high-resolution ($R\sim 40,000$) spectra were taken
from Keck on two successive nights (Castro et al.\ 2001).
Both initial papers focused on H$\alpha$.  Albrow et al.\ (2001b)
showed that the equivalent width (EW) of this line (which was unresolved)
varied during the four nights in the sense of being larger when
hotter parts of the stellar surface were more highly magnified, and
in particular that it dropped dramatically ($\sim 25\%$) on the last
night when only the extreme limb was highly magnified.  
Although Albrow et al.\ (2001b) do not mention it, such a sharp drop
would implies that the outer $\sim 4\%$ of the source is strongly in
{\it emission} in H$\alpha$, which would be in significant conflict with
the atmosphere model that they present.  Castro et al.\ 
(2001) measured an EW difference in H$\alpha$ between the two nights
of $8.3\pm 0.7\%$ and
showed that the optical depth difference is roughly constant over the
$\sim 15$ resolution elements that span the line.  See Figure \ref{fig:five}.
The declines in EW between July 6 and 7 measured by the two groups
are roughly consistent with one another, 
although the absolute normalization of the 
Albrow et al.\ (2001b) EW is about 10\% higher, perhaps reflecting blending
of H$\alpha$ 
in the low resolution spectrum with a line $\sim 1\,$\AA\ longward of
H$\alpha$ (see Fig.\ 2 of Castro et al.\ 2001).

	At present, it is not known exactly what can be learned about
stellar atmsopheres from studying microlensed spectra.  As mentioned above,
modeling has been mostly focused on point-mass caustic crossings, whereas
it is mostly binary caustics that will be observed spectroscopically.
Moreover, the theoretical studies carried out to date have not examined
whether different theoretical atmospheres predict detectably different
microlensed spectra.  EROS BLG-2000-5 presents a unique opportunity
for theorists to compare competing atmosphere models with two excellent data
sets to determine whether these distinguish between models.  Such 
a comparison would in turn give important clues as to how to carry
out observations of future caustic crossings.

\subsection{Microlenses as Telescopes}

	Microlensing events can be used simply as a method to amplify
the light gathering capabilities of one's telescope and so obtain
deeper spectra than would otherwise be possible.  This approach was
first applied by Bennetti, Pasquini \& West (1995) who were able to type
and measure the radial velocity ($-400\,\kms$) of a $V=20$ K0 subgiant using
3.6 m telescopes at low resolution.

	Although, Lennon et al.\ (1996) failed to resolve the source
(see \S 6.2),
they were able to use their magnification $\sim 25$ observations to measure 
the temperature and metallicity ([Fe/H]$\sim +0.3$) of a bulge dwarf star.

	Minnitti et al.\ (1998) obtained the first high-resolution 
($R=27,000$)
spectrum of a microlensing event, MACHO 97-BLG-45, and so by making use
of the high magnification were able to
obtain the first lithium abundance measurement for a bulge dwarf.

	In addition, for several microlensing events, spectra were taken
while the source was magnified in order to better characterize the
microlensing event itself (Albrow et al.\ 1998,1999a).

\subsection{Other Effects}

A number of other effects have been proposed that would probe
various aspects of the atmospheres of stars, although these have
not been met with the same level of interest from observers.

Gould (1997), inverting an idea of Maoz \& Gould (1994) showed that
spectra taken during a point-lens caustic crossing could be used
to measure rotation, even when the rotational broadening was far
smaller than the turbulent broadening.  

Simmons, Willis \& Newsam (1995)
demonstrated that stellar polarization could be measured during
a point-lens caustic crossing, even for a radially symmetric
polarization field.  Of course, in the absence of microlensing,
no such effects would be observable for an unresolved source.
A number of studies were then carried out for more complicated
geometries (Simmons, Newson \& Willis 1995; Algol 1996;
Belokurov \& Sazhin 1997)

Finally, Igance \& Hendry (1999), Han et al.\ (2000)
Heyrovsk\'y \& Sasselov (2000), and Bryce \& Hendry (2001),
have studied the detection of 
stellar spots when a source transits either a point-lens caustic or 
a fold caustic.

\section{Conclusions}

	Microlensing has emerged as a powerful
probe of stellar astrophysics.  The problem of how to use microlensing
to measure the stellar MF (including dark objects, BDs, WDs, NSs, BHs)
has been solved theoretically, and the practical instrument that can
make these measurements, {\it SIM}, is being built.  {\it SIM} can
also be used to make $\sim 1\%$ measurements of a few dozen
nearby stars, which would provide a precision test of stellar models.
Both of these prospects still lie almost a decade in the future, but
on other fronts, microlensing is already beginning to have an impact
on stellar astrophysics.  A significant number of binary events have
been observed and characterized, and these could already be used to
constrain the companion mass-ratio distribution for bulge stars.
To date, there have been four LD measurements
using microlensing, including one that was very precise, and another
that was of a metal-poor A star in another galaxy.  And very recently,
highly coordinated efforts of the microlensing community have produced
the first spatially-resolved spectroscopic measurements of a star other
than the Sun.  These are impressive accomplishments for a field that,
a decade ago, was thought of only in terms of studying dark matter.

{\bf Acknowledgements}:
I thank Scott Gaudi for many useful comments and criticisms of the
original manuscript. 
This work was supported in part by grant AST 97-27520 from the NSF and
in part by a grant from Le Minist\`ere de l'Education Nationale de la
Recherche et de la Technologie.

\clearpage 

\begin{figure}
\epsfxsize \hsize
\epsfbox{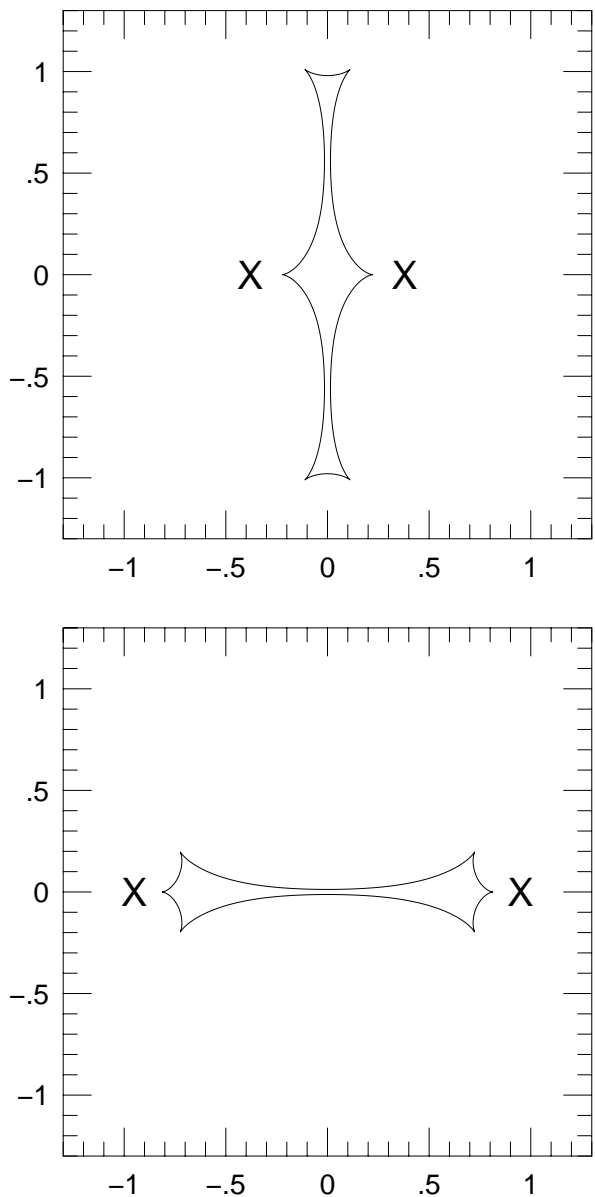}
\caption{Two extreme examples of 6-cusp caustics generated by
equal mass binaries.  The tick marks are in units of Einstein radii.
In each case, the crosses show the positions of the two components.
The upper panel shows a relatively close binary with the components
separated by $d=0.76$ Einstein radii.  For $d<2^{-1/2}$ the caustic would
break up into three caustics, a central 4-cusp caustic plus two outlying
3-cusp caustics.  The lower panel shows a relatively wide binary with $d=1.9$.
For $d>2$ the caustic would break up into two 4-cusp caustics.  
 From Gould (2001).  Copyright Astronomical Society of the Pacific, 
reproduced with permission.
}
\end{figure}

\begin{figure}
\epsfxsize=4.5in
\epsfbox{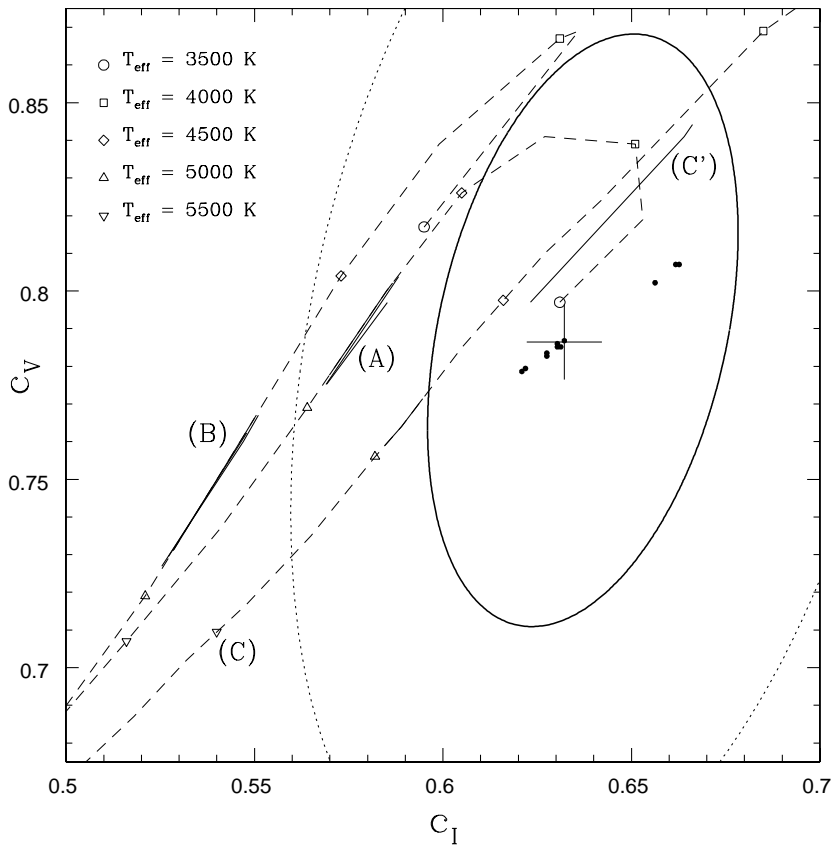}
\caption{\label{fig:four}
Comparison of linear limb-darkening coefficients in $V$ and $I$ derived
from stellar models and microlensing data for the G/K bulge subgiant in
OGLE 2000-BUL-23. The measured value from
the best model is represented by a small cross. One (\emph{solid line})
and two (\emph{dotted line}) $\sigma$ error ellipses
are also shown. Small dots are the results with
different global $(d,q)$ parameters. Various model predictions
are displayed by dashed lines ($\log\,g=\,3.5$). Model (A) is taken from
D\'{\i}az-Cordov\'{e}s et al.\ (1995)
and Claret et al.\ (1995), (B) is from van Hamme (1993), and (C) is from
Claret (1998).
In particular, the predicted values in the temperature range
that is consistent with the source color measurement 
($\Teff=\,[4820\pm 110]$ K for
$\log\,g=\,3.0$; $\Teff=\,[4830\pm 100]$ K for $\log\,g=\,3.5$; and
$\Teff=\,[4850\pm 100]$ K for $\log\,g=\,4.0$) are emphasized by thick
solid lines. Model (C') is by Claret (1998) for stars of
$\Teff=\,(4850\pm 100)$ K for $\log\,g=\,4.0$. Although the measured value
of the limb-darkening coefficients alone favors this model, the 
required young age would imply a disk rather than bulge source, which would
be inconsistent with the lens-source relative proper motion measured
for this event.  From Albrow et al.\ (2001a).  Copyright American Astronomical
Society, reproduced with permission.
}
\end{figure}

\begin{figure}\
\epsfxsize=5.0in
\epsfbox{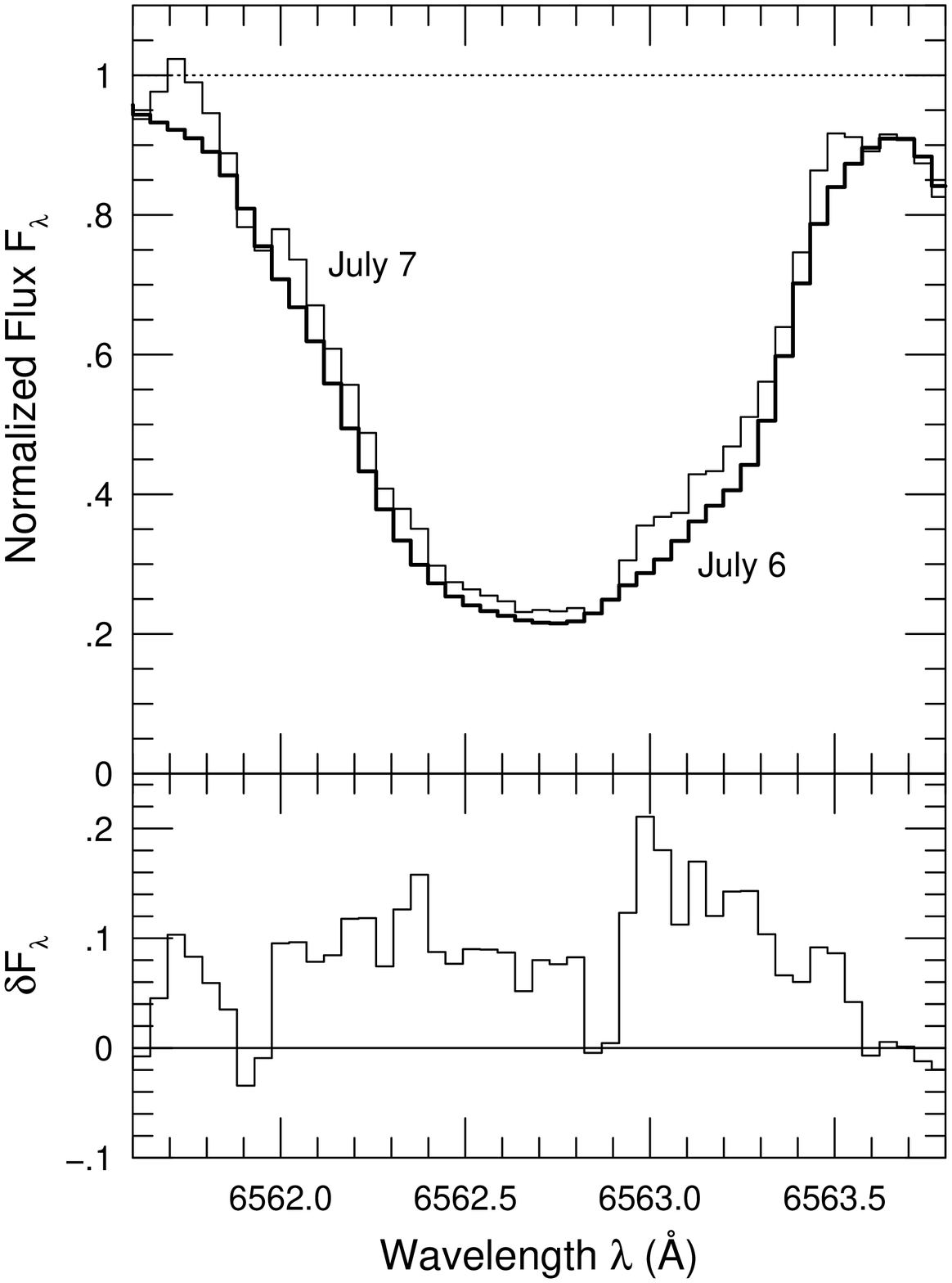}
\caption{\label{fig:five}
Upper panel: Keck HIRES spectra of a K3 bulge giant taken during the 
binary-lens 
caustic crossing of EROS BLG-2000-5 on the nights of 6 July ({\it bold}) 
and 7 July ({\it solid}) 2000.  Only the H$\alpha$ line at 
$\lambda=6562.7$\AA\ is shown.  
The full spectrum covers the range
$5500$\AA$<\lambda<7900$\AA.  The equivalent width of H$\alpha$ 
is $8.7\pm 0.7\%$ smaller on the second night. 
The spectra have each been normalized to a continuum of unity
and have been convolved to the same resolution.  Lower panel:  
Fractional difference in the lines between the two nights as a function of 
wavelength, $\delta F_\lambda \equiv 2(F_6 - F_7)/(F_6+F_7)$, where
$F_6$ and $F_7$ are the normalized fluxes from July 6 and July 7 respectively.
 From Castro et al.\ (2001).  Copyright American Astronomical
Society, reproduced with permission.
}
\end{figure}


\bigskip

\begin{references} 
\reference{} Afonso, C.\ et al.\ 2000, \apj, 532, 340
\reference{} Albrow, M.\ et al.\ 1998, \apj, 512, 672  
\reference{} Albrow, M.\ et al.\ 1999a, \apj, 522, 1011 
\reference{} Albrow, M.\ et al.\ 1999b, \apj, 522, 1022 
\reference{} Albrow, M.\ et al.\ 2000, \apj, 534, 894 
\reference{} Albrow, M.\ et al.\ 2001a, \apj, 549, 759 (astro-ph/0004243)
\reference{} Albrow, M.\ et al.\ 2001b, \apj\ 550, L173 
(astro-ph/0011380) 
\reference{} Alcock, C., et al.\ 1995, \apj, 454, L125 
\reference{} Alcock, C., et al.\ 1997a, \apj, 479, 119 
\reference{} Alcock, C., et al.\ 1997b, \apj, 491, 436 
\reference{} Alcock, C., et al.\ 1998, \apj, 499, L9 
\reference{} Alcock, C., et al.\ 2000a, \apj, 541, 270 
\reference{} Alcock, C., et al.\ 2000b, \apj, 541, 734 
\reference{} Alcock, C., et al.\ 2000c, \apj, 542, 281 
\reference{} Algol, E.\ 1996, \mnras, 279, 571
\reference{} Baruffolo, A., Benacchio, L., \& Benfante, L.\ 1999,
Astronomical Data Analysis Software and Systems VIII, ed. D.M. Mehringer,
ASP Conference Series, 172, 237 (San Francisco: ASP)
\reference{} Belokurov, V.A., \& Sazhin, M.V.\ 1997, Astronomy Reports, 41 777
\reference{} Benetti, S., Pasquini, L., \& West, R.M.\ 1995, \aap, 294, L37
\reference{} Bennett, D.P., et al.\ 1997, BAAS, 191, 8303
\reference{} Binney, J., Bissantz, N., \& Gerhard, O.\ 2000, \apj, 537, L99
\reference{} Boden, A.F., Shao, M., \& Van Buren, D.\ 1998 \apj, 502, 538
\reference{} Bogdanov, M.B., \& Cherepashchuk, A.M.\ 1995, ARep, 39, 779
\reference{} Boutreux, T., \& Gould, A.\ 1996, \apj, 462, 705
\reference{} Bryce, H.M., \& Hendry, M.A.\ 2001, in Microlensing 2000: A New 
Era in Microlensing Astrophysics, eds.\ J.W.\ Menzies \& P.D.\ Sackett, ASP 
Conf.\ Series, in press (astro-ph/0004250)
\reference{} Burns, D., et al.\ 1997, \mnras, 209, L11
\reference{} Castro, S.M., Pogge, R.W., Rich, R.M., DePoy, D.L., \& Gould, A.\
2001, \apj\ 548, L197
\reference{} Chang, K.\ 1981, The Two-Body Gravitational Lens Effect,
Ph.D.\ Thesis, (Hamburg: Univ.\ of Hamburg)
\reference{} Chang, K., \& Han, C.\ 1999, \apj, 525, 434
\reference{} Chang, K., \& Refsdal, S.\ 1979, Nature, 282, 561
\reference{} Chang, K., \& Refsdal, S.\ 1984, \aap, 130, 157
\reference{} Claret, A., D\'\i az-Cordov\'es, J., \& Gim\'enez, A. 1995, 
\aaps, 114, 247
\reference{} Claret, A.\ 1998, \aap, 335, 647
\reference{} Deeg, H.J., Garrido, R., \& Claret, A.\ 2001,
New Astronomy, in press (astro-ph/0012435)
\reference{} Derue, F.\ 1999, \aap, 351, 87
\reference{} D\'\i az-Cordov\'es, J., Claret, A., \& Gim\'enez, A.\ 1995, 
\aaps, 110, 329
\reference{} Di Stefano, R., \& Mao, S.\ 1996, \apj, 457, 93
\reference{} Dominik, M.\ 1998, \aap, 329, 361 
\reference{} Dominik, M.\ 1999, \aap, 349, 108 
\reference{} Dominik, M., \& Sahu, K.C.\ 2000, \apj, 534, 213
\reference{} Duquennoy, A., \& Mayor, M. 1991, \aap, 248, 485 
\reference{} Einstein, A.\ 1936, Science, 84, 506
\reference{}Froeschle, M., Mignard, F., \& Arenou, F.\ 1997, Proceedings of 
the ESA Symposium 'Hipparcos -- Venice '97',  p.\ 49, ESA SP-402
\reference{} Gaudi, B.S., \& Gould, A.\ 1997a, 477, 152
\reference{} Gaudi, B.S., \& Gould, A.\ 1997b, 483, 83
\reference{} Gaudi, B.S., \& Gould, A.\ 1999, 513, 619
\reference{} Gaudi, B.S., Naber, R.M., \& Sackett, P.D.\ 1998, \apj, 502, L33
\reference{} Gould, A.\ 1992, \apj, 392, 442
\reference{} Gould, A.\ 1994a, \apj, 421, L71 
\reference{} Gould, A.\ 1994b, \apj, 421, L75 
\reference{} Gould, A.\ 1994c, \apj\ Letters, submitted, (astro-ph/9408060)
\reference{} Gould, A.\ 1995, \apj, 441, L21 
\reference{} Gould, A.\ 1996, \pasp, 108, 465
\reference{} Gould, A.\ 1997, \apj, 483, 98 
\reference{} Gould, A.\ 2000a, \apj, 532, 936 
\reference{} Gould, A.\ 2000b, \apj, 535, 928 
\reference{} Gould, A.\ 2000c, \apj, 542, 785 
\reference{} Gould, A.\ 2001, in Microlensing 2000: A New Era in Microlensing 
Astrophysics, eds.\ J.W.\ Menzies \& P.D.\ Sackett, ASP Conf.\ Series, 
in press  (Astro-ph/0004042) 
\reference{} Gould, A., \& Andronov, N.\ 1999, \apj, 516, 236
\reference{} Gould, A., Bahcall, J.N.\ \& Flynn, C.\ 1997, ApJ, 482, 913
\reference{} Gould, A., \& Han, C.\ 2000, \apj, 538, 653
\reference{} Gould, A., \& Salim, S.\ 1999, \apj, 524, 794
\reference{} Gould, A., \& Welch, D.L.\ 1996, \apj, 464, 212
\reference{} Griest, K.\ et al.\ 1991, \apj, 372, L79
\reference{} Han, C.\ 1997, \apj, 484, 555 
\reference{} Han, C.\ 2001, \mnras, in press (astro-ph/0010557)
\reference{} Han, C., Chun, M.-S., \& Chang, K.\ 1999, \apj, 526, 405
\reference{} Han, C., Park, S.-H., Kim, H.-I., \& Chang, K.\ 2000, 
\mnras, 316, 665
\reference{} Han, C., \& Gould, A.\ 1996, ApJ, 467, 540 
\reference{} Han, C., \& Gould, A.\ 1997, \apj, 480, 196
\reference{} Han, C., \& Kim, H.-I.\ 2000, \apj, 528, 687
\reference{} Hardy, S.J., \& Walker, M.A.\ 1995, \mnras, 276, L79
\reference{} Henry, T.J., \& McCarthy, D.W.Jr.\ 1993, \aj, 106, 773
\reference{} Heyrovsk\'y, D., 2001, \apj, submitted
\reference{} Heyrovsk\'y, D., \& Loeb, A.\ 1997, \apj, 490, 38
\reference{} Heyrovsk\'y, D., \& Sasselov, D.\ 2000, \apj, 529, 69
\reference{} Heyrovsk\'y, D., Sasselov, D., \& Loeb, A.\ 1999, \apj,
submitted (preprint astro-ph/9902273)
\reference{} Hog, E., Novikov, I.D., \& Polnarev, A.G.\ 1995 \aap, 294, 287
\reference{} Honma, M.\ 1999, \apj, 517, L35
\reference{} Jha, S., Charbonneau, D., Garnavich, P.M., Sullivan, D.J.,
Sullivan, T., Brown, T.M., \& Tonry, J.L.\ 2000, \apj, 540, L45
\reference{} Ignace, R., \& Hendry, M.A.\ 1999, \aap, 341, 201
\reference{} Kiraga, M.\ \& Paczy\'nski, B.\ 1994, \apj, 430, 101
\reference{} Kuijken, K.\ 1997, \apj, 486, L19
\reference{} Lasserre, T.L., et al.\ 2000, \aap, 355, L39
\reference{} Lennon, D.J., Mao, S., Fuhrmann, K., \& Gehren, T.\ 1996, \apj, 
471, L23
\reference{} Loeb, A., \& Sasselov, D.\ 1995, \apj, 449, L33
\reference{} Luyten, W.J.\ 1979, New Luyten Catalogue of Stars with Proper 
Motions Larger than Two Tenths of an Arcsecond (NLTT) (Minneapolis: Univ.\ 
Minnesota Press)
\reference{} Luyten, W.J.\ 1980, New Luyten Catalogue of Stars with Proper 
Motions Larger than Two Tenths of an Arcsecond (NLTT) (Minneapolis: Univ.\ 
Minnesota Press)
\reference{} Mao, S.\ 1999, \aap, 350, L19
\reference{} Mao, S., \& Paczy\'nski, B.\ 1996, \apj, 473, 57
\reference{} Maoz, D., \& Gould, A.\ 1994, \apj, 425, L67
\reference{} Minniti, D., Vandehei, T., Cook, K. H., Griest, K., \& 
Alcock, C.\ 1998, \apj, 499, L175
\reference{} Miralda-Escud\'e, J.\ 1996, \apj, 470, L113
\reference{} Monet, D.\ 1998, BAAS, 193, 120.03
\reference{} Miyamoto, M., \& Yoshii, Y.\ 1995, \aj, 110, 1427
\reference{} Nemiroff, R.J.\ \& Wickramasinghe, W.A.D.T.\ 1994, \apj, 424, L21
\reference{} Orosz, J.A., \& Hauschildt, P.H.\ 2001, \aap, in press,
(astro-ph/0010114)
\reference{} Paczy\'nski, B.\ 1986, \apj, 304, 1
\reference{} Paczy\'nski, B.\ 1991, \apj, 371, L63
\reference{} Paczy\'nski, B.\ 1995, Acta Astronomica, 45, 345 
\reference{} Paczy\'nski, B.\ 1998, \apj, 494, L23
\reference{} Peale, S.J.\ 1998, \apj, 509, 177
\reference{} Peale, S.J.\ 1999, \apj, 524, L67 
\reference{} Popowski, P., et al.\ 2001, in Microlensing 2000: A New Era in 
Microlensing Astrophysics, eds.\ J.W.\ Menzies \& P.D.\ Sackett, ASP Conf.\ 
Series, in press
\reference{} Popper, D.M.\ 1984, \aj, 89, 132
\reference{} Refsdal, S.\ 1964, \mnras, 128, 295
\reference{} Refsdal, S.\ 1966, \mnras, 134, 315
\reference{} Rhie, S.H.\ 1997, \apj, 484, 63
\reference{} Rhie, S.H., \& Bennett, D.B.\ 1999, preprint (astro-ph/9912050)
\reference{} Salim, S., \& Gould, A.\ 2000, \apj, 539, 241
\reference{} Salim, S., \& Gould, A.\ 2001, in preparation
\reference{} Sahu, K.C., Chaney, E., Graham, J., Kane, S., \& Wieldt, D.\ 1998,
BAAS, 192.0701
\reference{} Schneider, P., Ehlers, J., \& Falco, E.E.\ 1992, 
Gravitational Lenses (Berlin: Springer)
\reference{} Schneider, P., \& Weiss, A. 1986, \aap, 164, 237
\reference{} Schneider, P., \& Weiss, A. 1987, \aap, 171, 49
\reference{}Simmons, J.F.L., Newsam, A.M., \& Willis, J.P.\ 1995, \mnras, 
276, 182
\reference{} Simmons, J.F.L., Willis, J.P., \&  Newsam, A.M.\ 1995, \aap, 
293, L46
\reference{} Soszy\'nski, I., et al.\ 2001, \apj, in press (astro-ph/0012144)
\reference{} Twigg, L.W., \& Rafert, J.B., 1980, \mnras, 193, 773
\reference{} Udalski, A., Szymanski, M., Kaluzny, J., Kubiak, M., Krzeminski, 
W., Mateo, M.,Preston, G.W., \& Paczynski, B.\ 1993, 
Acta Astronomica, 43, 289 
\reference{} Udalski, A., et al.\ 1994, Acta Astronomica 44, 165 
\reference{} Udalski, A., Kubiak, M. \& Szyma\'{n}ski, M. 1997,
Acta Astronomica, 47, 319
\reference{} Valls-Gabaud, D.\ 1995, in Large Scale Structure in the Universe, 
ed. J.P.\ M\"ucket, S. Gottl\"ober, \& V. M\"uller (Singapore: World 
Scientific), 326
\reference{} Valls-Gabaud, D.\ 1996, in IAU Symposium 173, Astrophysical
Applications of Graviational Lensing, eds.\ C.S.\ Kochanek \& J.N> Hewitt
(Dordrecht: Kluwer), 237
\reference{} Valls-Gabaud, D.\ 1998, \mnras, 294, 747
\reference{} van Belle, G.T.\ 1999, \pasp, 111, 1515
\reference{} van Hamme, W.\ 1993, \aj, 106, 2096
\reference{} Walker, M.A.\ 1995, \apj, 453, 37
\reference{} Wilson, R.E., \& Devinney, E.J.\ 1971, \apj, 166, 605
\reference{} Witt, H.J.\ 1990, \aap, 236, 311
\reference{} Witt, H.J.\ 1995, \apj, 449, 42
\reference{} Witt, H.J.\ \& Mao, S.\ 1994, \apj, 429, 66
\reference{} Zhao, H.\ S., Spergel, D.N., \& Rich, R.M.\ 1995, ApJ, 440, L13
\reference{} Zoccali, M., S., Cassisi, S., Frogel, J.A., Gould, A.,
Ortolani, S., Renzini, A.,  Rich, R.M. 1999, \&
Stephens, A.\ 2000, \apj, 530, 418


\end{references}
\end{document}